\documentstyle[12pt]{article}
\newfont{\ffont}{msym10}                        
\newcommand{\beq}{\begin{equation}}             
\newcommand{\eeq}{\end{equation}}               
\newcommand{\bqry}{\begin{eqnarray}}            
\newcommand{\eqry}{\end{eqnarray}}              
\newcommand{\bqryn}{\begin{eqnarray*}}          
\newcommand{\eqryn}{\end{eqnarray*}}            
\newcommand{\NL}{\nonumber \\}                  
\newcommand{\preprint}[1]{\begin{table}[t]      
            \begin{flushright}                  
            \begin{large}{#1}\end{large}        
            \end{flushright}                    
            \end{table}}                        
\newcommand{\PD}[2]                             
    {\frac{\partial^{#2}}{\partial #1^{#2}}}    
\catcode`\@=11 \@addtoreset{equation}{section}  
\renewcommand{\theequation}                     
         {\arabic{section}.\arabic{equation}}   
\begin{document}
\preprint{TAUP-2193-94 \\ }
\title{5D Generalized Inflationary Cosmology}
\author{\\ L. Burakovsky\thanks {Bitnet:BURAKOV@TAUNIVM.TAU.AC.IL} \
and L.P. Horwitz\thanks {Bitnet:HORWITZ@TAUNIVM.TAU.AC.IL. Also at
Department of Physics, Bar-Ilan University, Ramat-Gan, Israel  } \\ \ }
\date{School of Physics and Astronomy \\ Raymond and Beverly Sackler
Faculty of Exact Sciences \\ Tel-Aviv University,
Tel-Aviv 69978, Israel}
\maketitle
\begin{abstract}
We consider 5D Kaluza-Klein type cosmological model with the fifth
coordinate being a generalization of the invariant ``historical'' time
$\tau $ of the covariant theory of Horwitz and Piron. We distinguish
between vacuum-, off-shell matter-,  and on-shell matter-dominated eras
as the solutions of the corresponding 5D gravitational field equations,
and build an inflationary scenario according to which passage from the
off-shell matter-dominated era to the on-shell one occurs, probably as a
phase transition. We study the effect of this phase transition on the
expansion rate in both cases of local $O(4,1)$ and $O(3,2)$ invariance of
the extended $(x^\mu ,\tau )$ manifold and show that it does not change
in either case. The expansion of the model we consider is not adiabatic;
the thermodynamic entropy is a growing function of cosmic time for the
closed universe, and can be a growing function of historical time for the
open and the flat universe. A complete solution of the 5D gravitational
field equations is obtained for the on-shell matter-dominated universe.
The open and the closed universe are shown to tend asymptotically to the
standard 4D cosmological models, in contrast to the flat universe which
does not have the corresponding limit.
Finally, possible cosmological implications are briefly discussed.
\end{abstract}
\bigskip
\medskip
{\it Key words:} general relativity, cosmology, Kaluza-Klein models,
relativistic mechanics

PACS: 98.20.--d, 98.80.--k, 98.80.Cq
\bigskip
\section{Introduction}
The possibility that space-time has more than four dimensions has
received much attention regarding its cosmological aspects
[1--8]. Investigations have focused on attempts to explain why the
universe presently appears to have only four space-time dimensions if it
is, in fact, a dynamically evolving $(4+k)$-dimensional manifold ($k$
being the number of extra dimensions). It has been shown that solutions
to the $(4+k)$-dimensional Einstein equations exist, for which
four-dimensional space-time expands while the extra dimensions contract
or remain constant [4--8]. It has been pointed out that the extra
dimensions can produce large amounts of entropy during the contraction
process \cite{AG}, thus providing an alternative resolution to the
horizon and flatness problems \cite{Guth}, as compared to the usual
inflationary scenario. It has been also suggested that experimental
detection of the time variation of the fundamental constants could
provide strong evidence for the existence of extra dimensions \cite{Mar}.

In the present paper we study the 5D cosmological model of
Kaluza-Klein type, with the fifth coordinate being a generalization of
the universal parametric ``historical'' time $\tau $ discussed by, for
example, Stueckelberg \cite{Stu} and Horwitz and Piron \cite{HP}. It has
been shown that gauge invariance of the Stueckelberg-Schr\"{o}dinger
equation requires the addition of a fifth gauge field \cite{SHA}; this
result also follows from Feynman's approach to the foundations of gauge
theories in a manifestly covariant framework \cite{LHS}. The equations of
motion for such a gauge field are of second order in the five-dimensional
manifold $(x^\mu ,\tau ),$ with metric (4,1) or (3,2); i.e., on the level
of the gauge fields, the parametric ``historical'' time has entered a
five-dimensional manifold, much in the way that the Newtonian time $t$
enters the four-dimensional Minkowski manifold as a consequence of the
requirements of full gauge invariance of the Schr\"{o}dinger equation.
The canonical quantization of this $U(1)$ gauge theory has been carried
out by Shnerb and Horwitz \cite{SH}, where it is shown that the standard
Maxwell theory is recovered in a ``correlation'' limit.

The present paper is concerned with a 5D theory of gravitation with
$\tau $ as a fifth coordinate, which originated in an earlier work
\cite{therm}, in which we considered the thermodynamics of a relativistic
$N$-body system, taking account of the mass distribution in such a system
\cite{BH1,ind}. In \cite{therm} we incorporated two-body interactions, by
means of the direct action potential $V(q),$ where $q$ is an invariant
distance in the Minkowski space, taking the support of the two-body
correlations to be in a $O(2,1)$ invariant subregion of the full
spacelike region of relative coordinates. We then established the energy
conditions on matter in order that the Einstein equations possess a
singularity in terms of $V(q),$ and showed that, for a class of power-law
attractive potentials, $V(q)\sim q^n,\;0<n\leq 3,$ the energy conditions
for a singularity to occur can be violated only in the case of local
$O(3,2)$ invariance of the $(x^\mu ,\tau )$ manifold.

We have found that an off-shell ensemble at high temperatures is
characterized by the equation of state $p=(\Gamma -1)\rho ;\;\;$ $p,\rho
\propto T^{\Gamma /(\Gamma -1)},$ with $\Gamma $ being equal to 3/2 in
the case of local $O(3,2)$ invariance of the $(x^\mu ,\tau )$
manifold\footnote{The denotement $\sigma $ stands for the 55-component of
the local metric on the $(x^\mu ,\tau )$ manifold which is $g^{\alpha
\beta }=(+,-,-,-,\sigma ).$}
($\sigma =1)$ and 5/4 in the case of local $O(4,1)$ invariance ($\sigma =
-1),$ so that in the latter case $p=\rho /4,\;\;$ $p,\rho \propto T^5.$

Off-shell matter\footnote{We use the term `off-shell' to describe matter
with continuous mass distribution, i.e., non-point spectrum, as for
off-shell states occuring in the propagators of quantum field theory.}
with the equation of state $p,\rho \propto T^5$ was introduced into the
standard cosmological model in ref. \cite{fried}. It was shown that such
matter has energy density comparable with that of standard radiation
(with the equation of state $p,\rho \propto T^4)$ at temperature $\sim
10^{12}$ K, so that the possibility for a phase transition from the
off-shell sector to the on-shell one (with possible compactification of
the fifth dimension \cite{N}), at critical temperature $\sim 10^{12}$ K,
should be taken into account; for example, in the case of a Bose gas, by
the mechanism of a high-temperature Bose-Einstein condensation
\cite{BEc}.

As we show in the present paper, a 5D cosmological model of Kaluza-Klein
type permits derivation of vacuum-, off-shell matter-, and
on-shell matter-dominated eras as the solutions of the corresponding 5D
gravitational field equations. These solutions enable one to construct
an inflationary scenario (inflationary solutions arise in a
vacuum-dominated era) according to which, as the universe expands and
cools down, a phase transition from the off-shell sector to the on-shell
one occurs, probably at temperature $\sim 10^{12}$ K \cite{fried}. We
study its effect on the rate of expansion and show that in both cases
of $(\sigma =-1)\rightarrow (\sigma =0)$ and $(\sigma =1)\rightarrow
(\sigma =0)$ phase transition\footnote{We use $\sigma =0$ to describe the
standard (on-shell) 3+1 case.} the expansion rate does not change.

We show that the model we are discussing does not expand adiabatically.
For the closed universe the thermodynamic entropy is a growing function
of cosmic time; for the flat and the open universe it can be a growing
function of historical time. The open and the closed models will be shown
to go to the 4D standard cosmological models as the universe expands, in
contrast to the flat model which does not have the corresponding limit.

We remark that some previous discussions of these questions have been
made in the framework of 5D Kaluza-Klein theory
\cite{MV}--\cite{PLW}. Mann and Vincent \cite{MV} have shown that the
vacuum (Kaluza-Klein type) solutons of the five-dimensional field
equations give rise to an effective radiation density ($\rho =3p)$
connected with the extra dimension. Ponce de Leon and Wesson \cite{PLW}
have interpreted the sourceless solutions of the five-dimensional
Kaluza-Klein equations as those of the four-dimensional Einstein
equations with effective matter properties. We also note that
inflationary models based on the Kaluza-Klein framework have been
considered by Shafi and Wetterich \cite{SW}, and by Gr{\o }n \cite{G}
who has derived a complete cosmological scenario within the framework
of Wesson's gravitational theory with the rest mass
as a fifth coordinate \cite{W}.

\section{The line element}
Similarly to \cite{therm}, we take the fifth-dimension subspace to be
homogeneous and without coupling to the other coordinate,
i.e., a maximally symmetric subspace of the 5D
space \cite{Wei}. Then the 5D metric becomes \cite{Wei}
\beq
^{(5)}ds^2=g_{\alpha \beta }dx^\alpha dx^\beta =g_{\mu \nu }(x^\rho )
dx^\mu dx^\nu +g_{55}(x^\rho )d\tau ^2,
\eeq
$$\alpha ,\beta =0,1,2,3,5;\;\;\;\mu ,\nu ,\rho =0,1,2,3.$$ As shown in
Appendix, the 5D gravitational field equations
\beq
^{(5)}R_{\alpha \beta }=8\pi G\left( ^{(5)}T_{\alpha \beta }-\frac{1}{3}
g_{\alpha \beta }\;^{(5)}T^\lambda _\lambda \right) ,
\eeq
with the source term
\beq
^{(5)}T_{\alpha \beta }=\left( \;^{(4)}T_{\mu \nu },\;p_5\right) ,\;\;\;
p_5=\sigma \mu _K\kappa ,
\eeq
where $\kappa $ is the density of the generalized Hamiltonian per unit
comoving three-volume (actually associated with the density of the
variable mass) and $\mu _K$ is the mass potential in relativistic
ensemble \cite{HSP}, reduce to the 4D Einstein equations
\beq
^{(4)}R_{\mu \nu }=8\pi G\left( ^{(4)}T_{\mu \nu }-\frac{1}{2}
g_{\mu \nu }\;^{(4)}T^\rho _\rho \right)
\eeq
in the case of no curvature in $\tau $ direction.

As an example of the use of the metric (2.1) in cosmology, consider a
spacially flat cosmological model with the line element (in refs.
\cite{MV,We,PLW} a similar structure has been used)
\beq
ds^2=e^{\bar{\nu }}dt^2-e^{\bar{\omega }}(dx^2+dy^2+dz^2)-e^{\bar{\mu }}
d\tau ^2,
\eeq
where $\bar{\nu },\bar{\omega },\bar{\mu }$ are assumed (here) to be
functions of time alone. A particular solution is obtained for
$\bar{\nu }=0,$ $e^{\bar{\omega }}=t,$
$e^{\bar{\mu }}=t^{-1}.$ In this case the line element
\beq
ds^2=dt^2-t(dx^2+dy^2+dz^2)-t^{-1}d\tau ^2
\eeq
is similar to the cosmological model found by Chodos and Detweiler
\cite{CD}. They interpreted the fifth dimension geometrically in the
usual Kaluza-Klein sense \cite{KK}. The time coordinate of the line
element (2.6) is the proper time shown on standard clocks at rest in the
3D spacial hyperplane orthogonal to the time- and $\tau $-directions.
This will in the following be referred to as ``cosmic time''. In the
case that 4D space-time is filled with a medium in which these clocks are
at rest, the coordinate system is said to be ``comoving''. These are the
usual terms from ordinary 4D cosmology. The expansion factor of the model
(2.6) is $R(t)=t^{1/2}.$ This universe expands too slowly to solve the
horizon and flatness problems \cite{Guth}.

By the proper choice of $s,$ the line element (2.5) can be reduced to
\beq
ds^2=dt^2-e^\omega (dx^2+dy^2+dz^2)-e^\mu d\tau ^2.
\eeq
For this line element, the nonvanishing Christoffel symbols are
(henceforth the prime stands for derivative with respect to the 5D line
element, and the dot for derivative with respect to cosmic time)
\beq
\Gamma ^5_{05}=\frac{\dot{\mu }}{2},\;\;\;\Gamma ^0_{55}=\frac{1}{2}
\dot{(e^\mu )}.
\eeq
Therefore, the geodesic equations for $t$ and $\tau $ read (see Appendix)
\beq
t^{''}+\frac{1}{2}\dot{(e^\mu )}(\tau ^{'})^2=0,
\eeq
\beq
\tau ^{''}+\dot{\mu }t^{'}\tau ^{'}=0.
\eeq
The Lagrangian of a free comoving particle is
\beq
L=e^\mu (\tau ^{'})^2-(t^{'})^2.
\eeq
Since $\tau $ is a cyclic coordinate (it does not appear in the
Lagrangian), the conjugate momentum
\beq
p_\tau =\frac{\partial L}{\partial \tau ^{'}}=e^\mu \tau ^{'}
\eeq
is a constant of motion, giving
\beq
\frac{d\tau }{ds}=p_\tau e^{-\mu }.
\eeq
Inserting (2.13) into Eq. (2.9) gives
\beq
t^{''}=-\frac{p_\tau ^2}{2}e^{-\mu }\dot{\mu }=\frac{p_\tau ^2}{2}(e^{
-\mu })^{'}(t^{'})^{-1}.
\eeq
Integration leads to
\beq
\frac{ds}{dt}=\left( p_\tau ^2e^{-\mu }+C^2\right) ^{-1/2},
\eeq
where $C$ is an arbitrary constant. It then follows from (2.13),(2.15)
that
\beq
\frac{d\tau }{dt}=\frac{p_\tau e^{-\mu }}{(p_\tau ^2e^{-\mu }+C^2)^{1/2}
}.
\eeq

We now shall consider the following generalization of the line element
(2.5) which lets a spacial curvature be different from zero and permits
direct comparison with the Kaluza-Klein models\footnote{The 5D metric
(2.18) corresponds to local $O(4,1)$ invariance of an extended $(x^\mu ,
\tau )$ manifold. The choice of the metric in the form
\beq
ds^2=dt^2-\frac{R^2(t)}{(1+\frac{1}{4}kr^2)^2}(dx^2+dy^2+dz^2)+A^2(t)d
\tau ^2,
\eeq
corresponding to local $O(3,2)$ invariance of an $(x^\mu ,\tau )$
manifold, will lead essentially to the results of this work. Both cases
(2.17),(2.18) are treated simultaneously in the system of the field
equations (3.2)-(3.4).}:
\beq
ds^2=dt^2-\frac{R^2(t)}{(1+\frac{1}{4}kr^2)^2}(dx^2+dy^2+dz^2)-A^2(t)d
\tau ^2,
\eeq
where $r^2=x^2+y^2+z^2,$ and $k=0,\pm 1$ characterizes the spacial
curvature. Comparison with Eq. (2.7) shows that $e^\mu =A^2,$ so that,
expressed in terms of $A,$ Eqs. (2.13), (2.16) take on the form
\beq
\frac{ds}{d\tau }=\frac{A^2}{p_\tau },
\eeq
\beq
\frac{ds}{dt}=(p_\tau ^2/A^2+C^2)^{-1/2},
\eeq
\beq
\frac{dt}{d\tau }=\frac{(p_\tau ^2/A^2+C^2)^{1/2}}{p_\tau /A^2}.
\eeq
Since the standard case of no curvature in $\tau $ direction corresponds
to $A={\rm const},$ in this case $ds/d\tau ={\rm const},$ so that the
line element reduces (as seen in Eq. (2.17),(2.18); see also Appendix) to
\beq
d\tau ^2=dt^2-\frac{R^2(t)}{(1+\frac{1}{4}kr^2)^2}(dx^2+dy^2+dz^2),
\eeq
which in turn can be reduced to the standard 4D Robertson-Walker metric
\beq
d\tau ^2=dt^2-R^2(t)\left[ \frac{d\rho ^2}{1-k\rho ^2}+\rho ^2(d\theta
^2+\sin ^2\theta d\phi ^2)\right]
\eeq
with the help of the transformation \cite{Rin}
\beq
\rho =\frac{r}{1+\frac{1}{4}kr^2}.
\eeq
We see that, as $A\rightarrow {\rm const},$ the 5D universe with the line
element (2.17),(2.18) passes over to the standard 4D Robertson-Walker
universe\footnote{We remark that the universe with the Robertson-Walker
type metric $$ds^2=d\tau ^2-R^2(\tau )\left[ \frac{d\rho ^2}{1-k\rho ^2}+
\rho ^2\Big( d\theta ^2+\sin ^2\theta (-d\beta ^2+\cosh ^2\beta d\phi
^2)\Big)\right], $$
where $\rho ,\theta ,\beta ,\phi $ are the coordinates in the restricted
Minkowski space (RMS) \cite{therm}, as $R(\tau )\rightarrow \infty ,$
passes over to the standard 4D Robertson-Walker universe as well.
Details will be explained elsewhere \cite{RMS}.} (a phase transition to
the on-shell sector at $T\sim 10^{12}$ K probably taking place).

\section{The field equations}
We consider the following field equations for a 5D space-time filled with
a perfect fluid, permitting a non-vanishing cosmological constant
(henceforth we shall use the system of units in which $c=8\pi G=1):$
\beq
R_{\alpha \beta }-\frac{1}{2}g_{\alpha \beta }R+\Lambda g_{\alpha \beta }
=T_{\alpha \beta },
\eeq
with the source term (2.3). We note that the cosmological constant could
be disposed of by considering instead space-time with a vacuum fluid
which is a perfect fluid with the equation of state $p=-\rho .$

The field equations (3.1) with $T_{\alpha \beta }$ in the form (2.3)
reduce to the following system \cite{AG,MV}:
\beq
\frac{\dot{R}^2+k}{R^2}+\frac{\dot{R}\dot{A}}{RA}=\frac{1}{3}(\Lambda +
\rho ),
\eeq
\beq
\frac{2\ddot{R}}{R}+\frac{\dot{R}^2+k}{R^2}+\frac{2\dot{R}\dot{A}}{RA}+
\frac{\ddot{A}}{A}=\Lambda -p,
\eeq
\beq
\frac{\ddot{R}}{R}+\frac{\dot{R}^2+k}{R^2}=\frac{1}{3}(\Lambda +p_5).
\eeq
The usual 4D equations of the Friedmann model are obtained by setting $A=
{\rm const}$ in Eqs. (3.2),(3.3) and neglecting Eq. (3.4).
The energy-momentum conservation $T^{\alpha \beta }_{;\beta }=0$ implies
\cite{AG,MV}
\beq
\dot{\rho }+3(\rho +p)\frac{\dot{R}}{R}+(\rho -p_5)\frac{\dot{A}}{A}=0.
\eeq
For the initial stage of the evolution, when the universe is hot, we can
use \cite{therm}
\beq
p_5=\sigma p.
\eeq
This result, in fact, follows easily from the definition of the
five-dimensional energy-momentum tensor (also discussed in \cite{therm})
which is obtained by
the extension of the usual energy-momentum tensor
\beq
T^{\mu \nu }=(p+\rho )u^\mu u^\nu -pg^{\mu \nu },\;\;\;u^\rho u_\rho =1
\eeq
to a five-dimensional form:
\beq
^{(5)}T_{\alpha \beta }=\left(\;^{(5)}T_{\mu \nu },\;^{(5)}T_{55}\right);
\eeq
the requirement that the limiting case of the corresponding gravitational
theory (for zero curvature in the $\tau $ direction) coincides with the
Einstein equations results in the identification (see Appendix) $^{(5)}T_
{\mu \nu }=\;^{(4)}T_{\mu \nu }$ and $^{(5)}T_{55}=\sigma \mu _K\kappa. $
Expressions for $p$ and $\rho ,$ using the grand canonical
ensemble obtained by Horwitz, Schieve and Piron \cite{HSP} in their study
of manifestly covariant statistical mechanics, were found in \cite{ind}
in terms of confluent hypergeometric functions. For $T$ small, one finds
that $p,\rho \propto T^6,$ $\rho \simeq 5p,$ and, in fact, that $\mu _K
\kappa \propto T^7$ is negligible in comparison with $\rho .$ On the
other hand, for $T$ large, one finds \cite{therm} that in the case of
local $O(4,1)$ invariance of $(x^\mu ,\tau )$ manifold, $p,\rho \propto
T^5,$ $p\simeq \rho /4\simeq \mu _K\kappa ,$ while in the case of local
$O(3,2)$ invariance, $p,\rho \propto T^3,$ $p \simeq \rho /2\simeq
\mu _K\kappa .$ For high
temperature, it therefore follows that (as discussed in \cite{therm})
\beq
T^{\alpha \beta }=(p+\rho )u^\alpha u^\beta -pg^{\alpha \beta },\;\;\;
u^\lambda u_\lambda =1,
\eeq
so that, in the local rest frame,
$T_{\alpha \beta }={\rm diag}\;(\rho ,-p,-p,-p,\sigma p),$ and Eq. (3.6)
is justified. Moreover, for a perfect fluid, we use the equation of state
\beq
p=(\Gamma -1)\rho ,
\eeq
where $\Gamma $ is a constant. It then follows \cite{therm} that in the
cases of $\sigma =-1$ $(O(4,1))$ and $\sigma =1$ $(O(3,2)),$ $\Gamma $
is equal to 5/4 and 3/2, respectively, so that the tensor (3.9) is
traceless in either case. We note that the form (3.9) of a source term
for the 5D field equations has been used by Wesson \cite{W}.

In the case of local $O(4,1)$ invariance of an $(x^\mu ,\tau )$ manifold,
$\sigma =-1,$ $p_5=-p,$ and Eq. (3.5) integrates to
\beq
R^{3\Gamma }A^{2-\Gamma }\rho ={\rm const}.
\eeq
The expansion is not adiabatic. As $A\rightarrow {\rm const,}$ Eq. (3.11)
takes on the standard form
\beq
R^{3\Gamma }\rho ={\rm const}.
\eeq
Similarly, in the case of local $O(3,2)$ invariance of an $(x^\mu ,
\tau )$ manifold, $\sigma =1,$ $p_5=p,$ and Eq. (3.5) gives
\beq
R^{3\Gamma }A^\Gamma \rho ={\rm const},
\eeq
wich again reduces to the standard form (3.12) as $A\rightarrow {\rm
const}.$ Note that both Eqs. (3.11) and (3.13) give for dust matter with
$p\approx 0$ $(\Gamma =1)$ (this can also be obtained
directly from (3.5) with $p=p_5=0)$
\beq
R^3A\rho ={\rm const},
\eeq
which, as $A\rightarrow {\rm const},$ reduce to the standard result
\beq
R^3\rho ={\rm const}.
\eeq

\section{Solutions to the field equations}
The standard strategy for solving cosmological equations, i.e., to
exclude $\rho $ and $p$ from the equation for $R$ in terms of $\rho ,p,$
with the help of the equations of state and energy-momentum conservation:
$p=(\Gamma -1)\rho ,$ $\rho \propto R^{-n},\;n=3,4,$ does not work in our
case, since, in view of (3.11),(3.13), $\rho \propto R^{-p}A^{-q}$
and $A$ is a function of cosmic time. We therefore have to express $R$ in
terms of a parameter which is independent of $A.$ Such a parameter is a
cosmological constant $\Lambda .$

We shall suppose that $A$ is a slowly varying function of $t,$ so that
one can neglect the term $\ddot{A}/A$ in Eq. (3.3). Then, for $\sigma =
-1$ $(O(4,1)),$ one derives from (3.2)-(3.4), by the exclusion of $\rho ,
p$ and $p_5$ with the help of (3.6),(3.10), the equation
\beq
\frac{\Gamma +1}{2(2\Gamma -1)}\ddot{R}R+\dot{R}^2+k=\frac{\Gamma }{3(2
\Gamma -1)}\Lambda R^2,
\eeq
which for $\Gamma =5/4$ reduces to
\beq
\frac{3}{4}\ddot{R}R+\dot{R}^2+k=\frac{5}{18}\Lambda R^2.
\eeq
For $\sigma =1$ $(O(3,2))$ one similarly obtains
\beq
\frac{5\Gamma -3}{2\Gamma }\ddot{R}R+\dot{R}^2+k=\frac{2\Gamma -1}{3
\Gamma }\Lambda R^2,
\eeq
which for $\Gamma =3/2$ reduces to
\beq
\frac{3}{2}\ddot{R}R+\dot{R}^2+k=\frac{4}{9}\Lambda R^2.
\eeq
For the standard case of $p_5=0$ (or $\sigma =0)$ one gets
\beq
\ddot{R}R+\dot{R}^2+k=\frac{1}{3}\Lambda R^2,
\eeq
which really represents Eq. (3.4) with $p_5=0.$ Note that both Eqs. (4.1)
and (4.3) reduce to (4.5) for $\Gamma =1.$

\subsection{Vacuum-dominated era}
In a vacuum-dominated era the universe is filled with a vacuum fluid.
Eqs. (4.2),(4.4), \\ (4.5) can be represented by an equation of the
general form
\beq
a\ddot{R}R+\dot{R}^2+k=b\Lambda R^2,
\eeq
which, through the substitution
\beq
R^{1+\frac{1}{a}}=\tilde{R},
\eeq
reduces to the equation
\beq
\ddot{\tilde{R}}-\frac{b(a+1)}{a^2}\Lambda \tilde{R}+\frac{a+1}{a^2}k
\tilde{R}^{\frac{1-a}{1+a}}=0
\eeq
having the solution ($C_1,C_2={\rm const})$
\beq
\tilde{R}=C_1\cosh \sqrt{\frac{b(a+1)}{a^2}\Lambda }\;t+C_2\sinh \sqrt{
\frac{b(a+1)}{a^2}\Lambda }\;t+\left( \frac{k}{b\Lambda }\right) ^{\frac{
1}{2}(1+\frac{1}{a})}.
\eeq
We, therefore, obtain from (4.2),(4.4),(4.5), respectively: \\

for $\sigma =-1,$
\beq
R^{7/3}=\left( \frac{18}{5}\frac{k}{\Lambda }\right) ^{7/6}+C_1^{-}\cosh
\frac{\sqrt{70\Lambda }}{9}t+C_2^{-}\sinh \frac{\sqrt{70\Lambda }}{9}t,
\eeq

for $\sigma =1,$
\beq
R^{5/3}=\left( \frac{9}{4}\frac{k}{\Lambda }\right) ^{5/6}+C_1^{+}\cosh
\frac{\sqrt{40\Lambda }}{9}t+C_2^{+}\sinh \frac{\sqrt{40\Lambda }}{9}t,
\eeq

for $\sigma =0,$
\beq
R^2=\frac{3k}{\Lambda }+C_1^{0}\cosh \frac{\sqrt{54\Lambda }}{9}t+
C_2^{0}\sinh \frac{\sqrt{54\Lambda }}{9}t.
\eeq
We note that the solution (4.12) was obtained previously by Gr{\o }n
\cite{G}. In subsequent consideration we shall, for simplicity, restrict
ourselves to this solution alone. It then follows that, without any loss
of generality, this solution can be represented
by the following relations:

$k=1,$
\beq
R^2=\frac{2}{\omega ^2}(1+\cosh \omega t)=\frac{4}{\omega ^2}\cosh ^2
\frac{\omega t}{2},
\eeq

$k=-1,$
\beq
R^2=\frac{2}{\omega ^2}(\cosh \omega t-1)=\frac{4}{\omega ^2}\sinh ^2
\frac{\omega t}{2},
\eeq

$k=0,$
\beq
R^2=\frac{4}{\omega ^2}\exp\;(\omega t),
\;\;\;w\equiv \sqrt{\frac{2}{3}\Lambda }.
\eeq
Consider, for example, (4.13). For $t=0$ it gives $R=2/\omega .$ Since
the classical description of the expansion of the universe cannot be
valid prior to $t\sim t_{Pl}=M_{Pl}^{-1}\sim 5\cdot 10^{-44}$ s after the
big bang or the start of inflation at $t=0,$ one finds that $\omega
\stackrel{<}{\sim }M_{Pl}\sim 10^{19}$ GeV, and therefore $\Lambda =
\frac{3}{2}\omega ^2\stackrel{<}{\sim }10^{38}$ GeV$^2.$ By introducing
the vacuum energy density through the relation
\beq
\Lambda c^2=8\pi G\rho _{vac}
\eeq
and recovering $c$ and $G$ for numerical calculation, $G\sim M_{Pl}^{-2}
\simeq 10^{-38}$ GeV$^{-2},$ one obtains
\beq
\rho _{vac}<M_{Pl}^4/16\sim 10^{75}\;{\rm GeV}^4\simeq 10^{92}\;
{\rm g\;cm}^{-3}.
\eeq
If, similar to the standard inflationary models \cite{Linde}, one takes
$\rho _{vac}=T_c^4\sim 10^{60}$ GeV$^4,$ where $T_c\sim 10^{15}$ GeV is
a typical critical temperature for a phase transition in grand unified
theories \cite{Linde}, one obtains $\omega =\sqrt{16\pi G\rho _{vac}/3c^
2}\simeq 4\cdot 10^{11}$ GeV. As is usually done in the standard
inflationary models \cite{Linde}, inflation comes to an end when its rate
$H\equiv \dot{R}/R=\omega /2$ begins to decrease rapidly (which means
that the universe becomes rapidly increasing in size), the typical
time of inflation is $t_{inf}\sim 1/H=2/\omega .$ With $\omega \simeq 4
\cdot 10^{11}$ GeV, one finds $t_{inf}\simeq 10^{-36}$ s.

The value of the vacuum energy density (4.16) should be related to the
present-day vacuum energy density which is not much greater in absolute
value than the critical density $\rho _{cr}\sim 10^{-29}\;{\rm g\;cm}^{
-3},$ as implied by recent cosmological data. As remarked by Linde, in
grand unified theories (e.g., in the $SU(5)$ Coleman-Weinberg theory
\cite{CW}) this value of the vacuum energy density is obtained as a
result of a series of phase transitions. Other theories having the
cosmological constant (and, therefore, vacuum energy density) decreasing
with time are also discussed in the literature (e.g., a scale-covariant
theory of fundamental interactions \cite{Wes} in which $\Lambda \propto
t^{-2}).$ In ref. \cite{RSS} in which $A(t)$ is related to the quantum
one-loop correction terms as $\rho \sim -p_5\sim A^{-5},$ the problem is
surmounted by demanding that $A(t)$ be constant with a value cancelling
out the contribution from the 5D cosmological constant, producing a zero
effective 4D one.

As usually done in standard inflationary models \cite{Linde}, when
inflation ends, the cosmological constant $\Lambda $ is omitted in the
field equations (3.2)-(3.4), and subsequent evolution is described by a
Friedmann-type hot universe model. One may also think that the
cosmological constant is contained (through vacuum energy density) as
part of the off-shell matter energy density. This may have a reasonable
basis, since one notes that the off-shell matter energy density with
temperature dependence $\sim T^5$ \cite{ind} which at $T\sim 10^{28}$ K
$(=10^{15}$ GeV) is equal to $10^{75}$ GeV$^4$ can be consistently
represented by $$\rho ^{'}=10^{75}\left( \frac{T}{10^{28}}\right) ^5{\rm
GeV}^4;$$ it then follows from this formula that $\rho ^{'}$ takes on the
value $10^{-4}\;{\rm GeV}^4\sim 10^{14}\;{\rm g\;cm}^{-3},$ which is a
typical energy density of radiation-like matter at $T\sim 2\cdot 10^{12}
\;{\rm K}\simeq 150\;{\rm MeV},$ at the same temperature, implying the
possibility of a phase transition. Such a phase transition is briefly
analyzed in Section 5.

\subsection{Off-shell matter-dominated era}
In an off-shell matter-dominated era the universe is filled with an
off-shell fluid having the equation of state (3.10) with $$\Gamma =
\left\{ \begin{array}{ll}
5/4, & \sigma =-1, \\
3/2, & \sigma =1.
\end{array} \right. $$ In this case, as discussed below, we omit
$\Lambda $ in the field equations (3.2)-(3.4). Moreover, we can also omit
the spacial curvature $k$ which is negligible at high energy densities.
It then follows from Eq. (4.6) with zero r.h.s.
(this also follows from (4.9) for small $\sqrt{\Lambda }t)$ that
\beq
\tilde{R}=C_1^{'}+C_2^{'}t,
\eeq
so that

for $\sigma =-1,$
\beq
R^{7/3}=C_1^{'-}+C_2^{'-}t,
\eeq

for $\sigma =1,$
\beq
R^{5/3}=C_1^{'+}+C_2^{'+}t.
\eeq

\subsection{On-shell matter-dominated era}
In an on-shell matter-dominated era the universe is filled with a
standard on-shell radiation having the equation of state $p=1/3\rho ,$
and, as the universe expands and cools down, with dust matter with $p
\approx 0.$

For the universe filled with radiation, we omit both $\Lambda $ and
$k$ in the corresponding equation (4.5), which then has the solution
\beq
R^2=C_1^{'0}+C_2^{'0}t.
\eeq
Since Eqs. (4.19),(4.20) and (4.21) are obtained from (3.2)-(3.4) with
$p_5=\sigma p,$ and $p_5=0,$ respectively, the three Eqs. (4.19)-(4.21)
can be unified in one equation, as follows:
\beq
R^{2-\sigma \alpha /3}=C_1^{'\sigma }+C_2^{'\sigma }t,
\eeq
where
$$\alpha =\left\{ \begin{array}{ll}
1, & T\rightarrow \infty , \\
0, & T\rightarrow \;0,
\end{array} \right. $$ according to (see Eqs. (A.9),(A.11) of Appendix)
$$p_5=\left\{ \begin{array}{ll}
\sigma p, & T\rightarrow \infty , \\
0, & T\rightarrow \;0.
\end{array} \right. $$
For the universe filled with dust matter, we omit $\Lambda $ alone
in Eq. (4.5); this equation with zero r.h.s. has the solution
\beq
R^{2}=C_1^{''0}+C_2^{''0}t-kt^2.
\eeq

Two possible scenarios of evolution of the universe described by these
equations exist.

First, $\alpha $ in Eq. (4.22) is a smooth function of $T.$ As the
universe expands and its temperature decreases, $\alpha \rightarrow 0,$
so that in both cases $(\sigma =\pm 1)$ Eq. (4.22) passes over smoothly
to Eq. (4.21) which goes over to Eq. (4.23) at lower temperatures. That
is, the universe passes smoothly from an off-shell matter-dominated era
to on-shell one. In this case the rate of expansion is a smooth function
of temperature (and therefore the radius of the universe) as well.

Second, $\alpha $ is not a smooth function of $T,$ nor may it be a
function of $T$ at all. Since at some value of $R$ (and therefore $T)$
Eq. (4.19) (or (4.20)) goes over to Eq. (4.21), and the powers of $R$ in
the corresponding equations do not coincide, passage from an off-shell
matter-dominated era to on-shell one occurs as a {\it phase transition.}
In this case the rate of expansion does not change in either case of
$\sigma =-1$ or $\sigma =1,$ as we shall see below).

We consider the second scenario to be more realistic one, since a
passage from the off-shell sector (a sector of relativistic mass
distributions \cite{BH1,ind}) to on-shell one is probably a
phase transition\footnote{Another possibility is a smooth Galilean limit
$c\rightarrow \infty $ \cite{BH2}.}, as discussed in ref.
\cite{BEc} in the case of a relativistic Bose gas.

We now wish to discuss this phase transition in general features.

\section{Phase transition from off-shell matter-dominated era to on-shell
one}
Using Eqs. (4.19)-(4.21), we obtain the relations representing
continuity of $R$ at $t_0,$ where $t_0$ is the moment of cosmic time at
which the phase transition occurs:

$(\sigma =-1)\rightarrow (\sigma =0),$

\beq
(C_1^{'-}+C_2^{'-}t_0)^{3/7}=(C_1^{'0}+C_2^{'0}t_0)^{1/2},
\eeq

$(\sigma =1)\rightarrow (\sigma =0),$

\beq
(C_1^{'+}+C_2^{'+}t_0)^{3/5}=(C_1^{'0}+C_2^{'0}t_0)^{1/2}.
\eeq
We shall study the effect of the phase transition on the expansion rate.
We restrict our consideration to the case where the phase transition
occurs smoothly and adiabatically\footnote{More general cases of a
cosmological phase transition are considered in ref. \cite{GK} on the
example of a hadronic matter--the quark-gluon plasma phase transition.}.

It follows from Eqs. (3.2)-(3.4),(3.6),(3.10), through the exclusion of
$\Lambda ,$ that the following equations for $R$ in terms of $\rho $
hold,
\beq
\dot{R}^2=\frac{1}{2}\ddot{R}R+\rho \left[ \frac{\Gamma -1}{2}\left( 1+
\frac{\sigma }{3}\right) +\frac{1}{3}\right] -k,
\eeq
which reduces, in the corresponding cases, to:

for $\sigma =-1,\;\Gamma =5/4,$

\beq
\dot {R}^2=\frac{1}{2}\ddot{R}R+\frac{5}{12}\rho R^2-k,
\eeq

for $\sigma =1,\;\Gamma =3/2,$

\beq
\dot {R}^2=\frac{1}{2}\ddot{R}R+\frac{2}{3}\rho R^2-k,
\eeq

for $\sigma =0,\;\Gamma =4/3$ (standard case),

\beq
\dot {R}^2=\frac{1}{2}\ddot{R}R+\frac{1}{2}\rho R^2-k.
\eeq
A smooth transition occurs at the constant pressure.
Using the corresponding equations of state $\rho =p/(\Gamma -1)$ and
Eqs. (5.4)-(5.6) (in which we neglect $k),$ we find the following
relations which represent equality of pressure in the corresponding
phases (one simply equates the quantities $(\Gamma -1)\rho R^2(t_0)):$

$(\sigma =-1)\rightarrow (\sigma =0),$

\beq
\frac{9}{49}\left( C_2^{'-}\right) ^2(C_1^{'-}+C_2^{'-}t_0)^{-8/7}=
\frac{1}{4}\left( C_2^{'0}\right) ^2(C_1^{'0}+C_2^{'0}t_0)^{-1},
\eeq

$(\sigma =1)\rightarrow (\sigma =0),$

\beq
\frac{9}{25}\left( C_2^{'+}\right) ^2(C_1^{'+}+C_2^{'+}t_0)^{-4/5}=
\frac{1}{4}\left( C_2^{'0}\right) ^2(C_1^{'0}+C_2^{'0}t_0)^{-1}.
\eeq
Calculation of $\dot {R}(t_0)$ gives, respectively,

for $\sigma =-1,$

\beq
\frac{3}{7}C_2^{'-}(C_1^{'-}+C_2^{'-}t_0)^{-4/7},
\eeq

for $\sigma =1,$

\beq
\frac{3}{5}C_2^{'+}(C_1^{'+}+C_2^{'+}t_0)^{-2/5},
\eeq

for $\sigma =0,$

\beq
\frac{1}{2}C_2^{'0}(C_1^{'0}+C_2^{'0}t_0)^{-1/2}.
\eeq
Comparison of Eqs. (5.9),(5.10) (squared) with Eq. (5.11) (squared),
using Eqs. (5.7) and (5.8), shows that $\dot {R}(t_0)$ in both $\sigma =-
1$ and $\sigma =1$ phases coincide with $\dot{R}(t_0)$ in the $\sigma =0$
phase; since $R(t_0)$ is the same for the three, we conclude that the
rate of expansion,
\beq
H\equiv \frac{\dot {R}}{R},
\eeq
{\it does not change} in either case of the $(\sigma =-1)\rightarrow
(\sigma =0)$ or $(\sigma =1)\rightarrow (\sigma =0)$ phase transitions.
This observation suggests that the phase transition should be
sufficiently smooth (second order). Although a first order phase
transition might be preferable for cosmological implications, due to the
fluctuations which are generated at the transition\footnote{The
fluctuations could not directly affect galaxy formation, since the
horizon size at the time of the transition is on a planetary scale
\cite{hor}. It has been demonstrated \cite{CS} that they could produce
planetary mass black holes; these black holes could provide a possible
explanation for the dark matter of the universe and even be seeds in
galaxy formation \cite{FPS,Wi}.}, experimental indications on the order
of this phase transition are still absent. Indeed, cosmological phase
transition at $T_c\sim 150$ MeV is normally associated with the
transition from a strongly interacting hadronic phase to a weakly
interacting quark-gluon plasma phase \cite{GK,OW}. Presently
available lattice data on $SU(N)$ pure gauge theory lattice simulations
indicate that a phase transition to a weakly interacting phase is of
apparently first order for $SU(3)$ and second order for $SU(2)$ theory
\cite{Mul}. In ref. \cite{Br}, however, it is argued that the apparent
first order nature of the transition in the case of $SU(3)$ pure gauge
theory may well be a lattice artefact. Moreover, there are indications
from lattice QCD calculations that when fermions are included, the phase
transition may be of second or higher order \cite{Ben}. In this case, as
remarked by Ornik and Weiner \cite{OW}, the phase transition would be
hardly distinguishable from a situation in which no phase transition
would have taken place (radiation-dominated universe alone).

\section{``Generalized'' entropy and the behavior of $A$ in an on-shell
matter-dominated era}
As we have seen in Section 3, the expansion of the universe with the line
element (2.17) is not adiabatic, due to the presence of time-dependent
$A$ in the equation (3.5) for energy-momentum conservation. Rewriting
this equation in the form
\beq
\left[ R^3A\rho \right] ^\bullet +Ap\dot {R^3}-p_5R^3\dot{A}=0,
\eeq
we see that the universe can be characterized by the ``first law''
\beq
d(AE)=ATd\tilde{S}-ApdV+p_5VdA,
\eeq
so that, in view of (6.1),(6.2) and $V\sim R^3,$
the ``generalized'' entropy $\tilde{S}$ is conserved:
\beq
\frac{d\tilde{S}}{dt}=0.
\eeq
It follows from (6.2) and genuine first law
\beq
dE=TdS-pdV
\eeq
that the thermodynamic entropy are related to the ``generalized'' one
as follows:
\beq
dS=d\tilde{S}-\frac{E^{'}}{T}\frac{dA}{A},
\eeq
where $E^{'}\equiv E-p_5V=\rho ^{'}V,$ and $$\rho ^{'}\equiv \rho -p_5$$
is the ``reduced'' energy density \cite{therm}. Hence, in view of (6.3),
\beq
\frac{dS}{dt}=-\frac{E^{'}}{T}\frac{1}{A}\frac{dA}{dt}.
\eeq
One sees that the sign of $\frac{dS}{dt}$ is determined by the sign of
$-\frac{1}{A}\frac{dA}{dt}.$ Note that, as $A\rightarrow {\rm const},$
it follows from (6.5) that $S=\tilde{S}+{\rm const}.$

In general, the time dependence of $A$ can be derived from Eqs.
(3.2)-(3.4), (3.6), (3.10), provided that the corresponding time
dependence of $R$ is known. We shall restrict ourselves to the on-shell
matter-dominated universe (similar consideration for the vacuum- and the
off-shell matter-dominated eras does not seem to present a difficulty).
For the universe filled with radiation-like matter, Eqs. (3.2)-(3.4) take
on the form
\beq
\frac{\dot{R}^2}{R^2}+\frac{\dot{R}\dot{A}}{RA}=\frac{1}{3}\rho =p,
\eeq
\beq
\frac{2\ddot{R}}{R}+\frac{\dot{R}^2}{R^2}+\frac{2\dot{R}\dot{A}}{RA}+
\frac{\ddot{A}}{A}=-p,
\eeq
\beq
\frac{\ddot{R}}{R}+\frac{\dot{R}^2}{R^2}=0.
\eeq
It follows from Eq. (6.9) that $$R^2=C^{'0}_1+C^{'0}_2t,$$ in agreement
with (4.21). Summing up Eqs. (6.7) and (6.8), taking into account (6.9),
gives
\beq
\frac{3\dot{R}\dot{A}}{RA}+\frac{\ddot{A}}{A}=0.
\eeq
This equation has the solution
\beq
\dot{A}=\frac{C}{R^3},\;\;\;C={\rm const};
\eeq
hence
\bqry
A & = & C\int \frac{dt}{R^3}\;=\;C\int \frac{dt}{(C^{'0}_1+C^{'0}_2t)^
{3/2}} \NL
 & = & \frac{-2C/C^{'0}_2}{(C^{'0}_1+C^{'0}_2t)^{1/2}}+\beta \;=\;
\frac{\alpha }{R}+\beta ,\;\;\;\alpha ,\beta ={\rm const}.
\eqry
Therefore
\beq
-\frac{1}{A}\frac{dA}{dt}=\frac{\dot{R}}{R(1+\frac{\beta }{\alpha }R)}.
\eeq
One sees that, since $\dot{R}>0,$ if $\alpha $ and $\beta $ are of the
same sign, or if $\beta =0,$ then $-\frac{1}{A}\frac{dA}{dt}>0,$ and
therefore $\frac{dS}{dt}>0,$ in view of (6.6).

For the dust universe, it follows from Eqs. (3.3),(3.4) with zero r.h.s.
that
\beq
\frac{\ddot{R}}{R}+\frac{2\dot{R}\dot{A}}{RA}+\frac{\ddot{A}}{A}=0,
\eeq
which reduces to the equation
\beq
\ddot{(AR)}=0,
\eeq
which has the solution (taking into account (4.23))
\beq
A=\frac{a+bt}{R}=\frac{a+bt}{\sqrt{C^{''0}_1+C^{''0}_2t-kt^2}},\;\;\;
a,b={\rm const}.
\eeq
The substitution
\beq
\rho =\frac{3\gamma }{AR^3},\;\;\gamma ={\rm const}
\eeq
(which follows from (3.14)) in the r.h.s. of Eq. (3.2) (in which we
omit $\Lambda ,$ as usual) yields, with the help of Eq. (3.4) with
zero r.h.s.,
\beq
\dot{R}\dot{A}-A\ddot{R}=\frac{\gamma }{R^2}.
\eeq
One can then find that Eqs. (6.16) and (6.18) are compatible if
\beq
\left\{ \begin{array}{ccc}
k & = & 0, \\
bC^{''0}_2 & = & 2\gamma ,
\end{array} \right.
\eeq
\beq
\left\{ \begin{array}{ccc}
k & \neq & 0, \\
a & = & \gamma k, \\
bC^{''0}_2 & = & 0,
\end{array} \right.
\eeq
and the same relations (6.19),(6.20) with $C^{''0}_1=0.$ Moreover, the
solutions (4.23) for $R$ and (6.16) for $A$ should be matched\footnote{By
matching we mean continuity of a function and its first derivative.} with
the corresponding solutions (4.21),(6.12) for the radiation-like
universe, which we rewrite here:
\beq
\left\{ \begin{array}{ccl}
R & = & \sqrt{C^{'0}_1+C^{'0}_2t}, \\
A & = & \frac{\alpha }{\sqrt{C^{'0}_1+C^{'0}_2t}}+\beta .
\end{array} \right.
\eeq
Thus, one is left with the following solutions for the dust universe:

for $k=0,$

\beq
\left\{ \begin{array}{ccl}
R & = & \sqrt{C^{''0}_1+C^{''0}_2t}, \\
A & = & \frac{a+2\gamma t/C^{''0}_2}{\sqrt{C^{''0}_1+C^{''0}_2t}},
\end{array} \right.
\eeq

for $k=1,$

\beq
\left\{ \begin{array}{ccl}
R & = & \sqrt{C^{''0}_1+C^{''0}_2t-t^2}, \\
A & = & \frac{\gamma }{\sqrt{C^{''0}_1+C^{''0}_2t-t^2}},
\end{array} \right.
\eeq

for $k=-1,$

\beq
\left\{ \begin{array}{ccl}
R & = & \sqrt{C^{''0}_1+t^2}, \\
A & = & \frac{bt-\gamma }{\sqrt{C^{''0}_1+t^2}}.
\end{array} \right.
\eeq

\subsection{Flat dust universe}
Consider first the case of the flat dust universe. It follows from (6.22)
that, as $t\rightarrow \infty ,$
\beq
A\sim R\sim t^{1/2}.
\eeq
Therefore, $-\frac{1}{A}\frac{dA}{dt}=-\frac{1}{R}\frac{dR}{dt}=-H(t)<0,$
so that the entropy is a decreasing function of cosmic time. In this
case, as seen in Eq. (2.21), if $p_\tau <0,$
\beq
\frac{dS}{d\tau }=\frac{dt}{d\tau }\frac{dS}{dt}>0,
\eeq
i.e., the entropy is a growing function of {\it historical
time}; if we moreover take $\alpha $ and $\beta $ in (6.12) of the
opposite sign, we will have, in view of (6.13), $\frac{dS}{d\tau }>0$
during all the on-shell matter-dominated era.

It is seen in Eq. (6.25) that for the model (6.22), the limit
$A\rightarrow {\rm const}$ is absent, i.e., it does not go over to the
standard 4D cosmological flat model, but rather represents the model
(2.5) with $e^{\bar{\omega }}=e^{\bar{\mu }}=t.$

\subsection{Closed dust universe}
Now we turn to the case of the closed dust universe. As seen in Eqs.
(6.23), the universe expands until the moment
\beq
t_0=\frac{C^{''0}_2}{2},
\eeq
reaching the maximal radius, and then begins to contract. For this model,
\beq
\frac{dA}{dt}=\frac{\gamma (t-\frac{C^{''0}_2}{2})}{(C^{''0}_1+C^{''0}_
2t-t^2)^{3/2}}=\frac{\gamma (t-\frac{C^{''0}_2}{2})}{R^3},
\eeq
so that, as $t\rightarrow t_0,$ $\frac{dA}{dt}\simeq 0,$ i.e., $A\simeq
{\rm const}.$ Therefore, in this limit the line element (2.17) goes over
to the standard 4D one, (2.23), and the 5D gravitational field equations
become indistinguishable from the standard 4D Einstein equations.

Since $A=\gamma /R,$
\beq
-\frac{1}{A}\frac{dA}{dt}=\frac{1}{R}\frac{dR}{dt}=H(t)=\frac{\frac{C^{''
0}_2}{2}-t}{C^{''0}_1+C^{''0}_2t-t^2}=\frac{\frac{C^{''0}_2}{2}-t}{R^2}.
\eeq
We see that, as $t\leq C^{''0}_2/2,$ $-\frac{1}{A}\frac{dA}{dt}\geq 0,$
and therefore $\frac{dS}{dt}\geq 0,$ via (6.6); hence, by virtue of
(6.14),(6.28), for the closed universe the entropy increases during the
expansion in the whole on-shell matter-dominated era, reaching its
maximum at $t=t_0,$ where $\frac{dS}{dt}=0.$

Rewriting (6.23) in the form
\beq
\left\{ \begin{array}{ccl}
R & = & \sqrt{C^{''0}_1+(C^{''0}_2)^2/4-\tilde{t}^2}, \\
A & = & \gamma /R,
\end{array} \right.
\eeq
where
\beq
\tilde{t}\equiv C^{''0}_2/2-t,
\eeq
we see that the model possesses explicit $\tilde{t}$-reversal. Since
\beq
\frac{dS}{d\tilde{t}}=-\frac{dS}{dt},
\eeq
one sees that, for $t>t_0,$ when the universe contracts, the entropy is a
growing function of $\tilde{t},$ as seen in Eqs. (6.29),(6.32).

Let us write down the formula which is valid for the closed universe and
follows from Eq. (6.6) (with $E^{'}=E$ for the on-shell matter-dominated
universe) and $A\sim 1/R:$
\beq
\frac{dS}{dt}=\frac{E}{T}H(t).
\eeq
Note also that for $A\sim 1/R,$ the energy-momentum conservation (3.5)
yields the relation (3.15) for the dust universe. Indeed, in the case of
local $O(4,1)$ invariance of the $(x^\mu ,\tau )$ manifold, it follows
from (3.11) with $A\sim 1/R$ that $R^{4\Gamma -2}\rho ={\rm const;}$ for
$\Gamma =5/4$ the latter reduces to (3.15). Similarly, in the $O(3,2)$
case, one obtains from (3.13) $R^{2\Gamma }\rho ={\rm const,}$ which
again reduces to (3.15) for $\Gamma =3/2.$

\subsection{Open dust universe}
For the open dust universe, as seen in Eqs. (6.24), as
$t\rightarrow \infty ,$
\beq
R\simeq t,\;\;\;A\simeq b-\gamma /t\rightarrow b={\rm const},
\eeq
so that this model tends asymptotically to the standard 4D cosmological
open model, for which $R\simeq t$ at large $t$ \cite{Miln}. It follows
from (6.24) that, as $t\rightarrow \infty ,$
\beq
-\frac{1}{A}\frac{dA}{dt}=-\frac{bC^{''0}_1+\gamma t}{(C^{''0}_1+t^2)^{3/
2}}\rightarrow -\frac{\gamma }{t^2}<0,
\eeq
since $A$ and $\gamma $ are of the same sign, in view of (6.17). Thus,
for the open dust universe, the entropy is a decreasing function of
cosmic time, but it can be a growing function of historical time, if $p_
\tau <0,$ similarly to the case of the flat dust universe.

\section{Concluding remarks}
We have considered 5D Kaluza-Klein type cosmological model with the fifth
coordinate being an invariant historical time $\tau .$ We have derived a
complete cosmological inflationary scenario for such a model which
distinguishes between vacuum-, off-shell matter-, and on-shell
matter-dominated eras as the solutions of the corresponding 5D
gravitational field equations. According to this scenario, the passage
from the off-shell matter-dominated era to the on-shell one occurs,
probably as a phase transition. We have studied the effect of this phase
transition on the expansion rate and found that it does not change in
either case of local $O(4,1)$- or $O(3,2)$-invariance of the extended
$(x^\mu ,\tau )$-manifold.

In contrast to the standard cosmological model in which the expansion of
the universe is adiabatic, $dS/dt=0$ \cite{fried,N}, the model considered
here does not expand adiabatically; the thermodynamic entropy is a
growing function of cosmic time for the closed universe, and can be a
growing function of historical time for the open and the flat universes.

We have obtained a complete solution of the 5D gravitational field
equations for the on-shell matter-dominated universe. We have shown that
the 5D open and closed universes tend asymptotically to the
corresponding standard 4D cosmological models, in contrast to the 5D flat
universe which does not have such a corresponding limit.

The question of the choice of the source term in the form containing $p_
5$ (like (2.3)) has received attention in the recent literature.
Mann and Vincent \cite{MV} have considered the case of local $O(4,1)$
invariance of 5D manifold and used the source term with $p_5$ involved.
The effective 4D equation of state obtained by them from the vacuum
solution of the 5D equations, $\rho =3p,\;p_5=0,$ is essentially the one
used in ref. \cite{RSS} with the quantum one-loop correction terms ($\rho
\sim -p_5\sim A^{-5})$ neglected in comparison with the classical 5D
radiation term. Wesson \cite{W} has considered the 5D version of 4D field
equations, $R_{\alpha \beta }=T_{\alpha \beta },$ with the source term
$T_{\alpha \beta }$ in the form (3.9) but only solved $R_{\alpha \beta }=
0.$ Later on, Wesson and independently Ponce de Leon \cite{WPL}
suggested that the 5D field equations may be just $R_{\alpha \beta }=0,$
and the extra terms which appears on the left-hand sides of the 5D
equations $R_{\alpha \beta }=0$ may correspond to the terms involving
matter parameters (like the density and pressure) which appear on the
right-hand sides of the 4D equations $R_{\mu \nu }=T_{\mu \nu }.$ More
recently, Ponce de Leon and Wesson \cite{PLW} have shown that a 5D theory
with no source can be cast into the form of a 4D theory with a source, in
the three-dimensionally symmetric case. In this case (when the metric is
independent of the extra coordinate, which is the case we consider in our
paper), as they have shown, the equation of state has the form of that of
radiation-like matter, $p=\rho /3.$ Hence, it is not possible, in the
case when the source term is obtained from the geometry of the higher
dimensional theory alone, to achieve the equation of state of a strongly
interacting phase (e.g., $p=\rho /4,$ which has certain experimental
evidence \cite{exp}), as we have obtained with an explicit source term,
nor to describe the inflationary epoch.

By rewriting the equation of the energy-momentum conservation, (3.5), in
the form $$\dot{\rho }+\left( 3\frac{\dot{R}}{R}+\frac{\dot{A}}{A}\right)
(\rho +p)-\frac{\dot{A}}{A}(p+p_5)=0,$$ one sees that the question of the
presence of $p_5$ in the source term for the field equations (3.2)-(3.4)
is associated with the question of whether or not the cosmological fluid
is ideal. In the case of local $O(4,1)$ invariance of $(x^\mu ,\tau )$
manifold, when $p_5=-p,$ the latter equation reduces to $$\dot{\rho }+
3\frac{\dot{R^{'}}}{R^{'}}(\rho +p)=0,$$ which is the standard equation
for ideal cosmological fluid expanding non-adiabatically due to the
scale factor $R^{'}=RA^{1/3}.$ In the other cases, $p_5=p$ $(O(3,2))$ or
$p_5=0$ \cite{MV}, the cosmological fluid is not ideal since the
energy-momentum tensor has an anisotropic pressure. Note that, as follows
from (3.5), in the case of local $O(3,2)$ invariance of $(x^\mu ,\tau )$
manifold, the cosmological fluid is ideal and expands $adiabatically$ if
$p_5=\rho ,$ implying the equality of the energy and mass densities, as
for standard radiation-like matter. Evolution of the universe filled with
such a fluid will be the subject of subsequent study.

\newpage
\appendix
\section*{Appendix}
\setcounter{equation}{0}
\renewcommand{\theequation}{A.\arabic{equation}}
In this Appendix we review briefly a five-dimensional theory of
gravitation with $\tau $ as a fifth coordinate. Full consideration will
be given elsewhere \cite{cosm}. We shall proceed in a way similar to the
one chosen by Ma \cite{Ma} within a different framework. Suppose that the
fifth-dimension subspace is homogeneous and without coupling to the other
coordinate, i.e., it is a maximally symmetric subspace of the 5D space
\cite{Wei}. With this assumption, the 5D metric becomes \cite{Wei}
\beq
^{(5)}ds^2=g_{\alpha \beta }dx^\alpha dx^\beta =g_{\mu \nu }(x^\rho )
dx^\mu dx^\nu +g_{55}(x^\rho )d\tau ^2,
\eeq
$$\alpha ,\beta =0,1,2,3,5;\;\;\;\mu ,\nu ,\rho =0,1,2,3.$$ The
line element in the hypersurface $\tau ={\rm const}$ is that of the 4D
space-time: $$^{(4)}ds^2=g_{\mu \nu }(x^\rho )dx^\mu dx^\nu .$$ Since
$g_{\alpha \beta }g^{\alpha \gamma }=\delta _\beta ^\gamma $ implies
$g_{\mu \nu }g^{\mu \rho }=\delta _\nu ^\rho $ and $g_{55}g^{55}=1,$
the only non-zero components of the 5D Christoffel symbol $^{(5)}
\Gamma ^\alpha _{\beta \gamma }$ are
\beq
^{(5)}\Gamma ^\mu _{\nu \rho }=\; ^{(4)}\Gamma ^\mu _{\nu \rho },\;\;\;
^{(5)}\Gamma ^5_{5\nu }=\frac{1}{2}g_{55}\frac{\partial g_{55}}{
\partial x^\nu },\;\;\; ^{(5)}\Gamma ^\mu _{55}=-\frac{1}{2}g^{\mu
\nu }\frac{\partial g_{55}}{\partial x^\nu }.
\eeq
The non-vanishing components of the 5D Ricci tensor and the 5D scalar
curvature are, respectively,
\bqry
^{(5)}R_{\mu \nu } & = &  ^{(4)}R_{\mu \nu }+\frac{\partial \Gamma ^5_
{\mu 5}}{\partial x^\nu }+\Gamma ^5_{\mu 5}\Gamma ^5_{\nu 5}-\Gamma ^
\rho _{\mu \nu }\Gamma ^5_{\rho 5}, \NL
^{(5)}R_{55} & = & -\frac{\partial \Gamma ^\mu _
{55}}{\partial x^\mu }+\Gamma ^\mu _{55}\Gamma ^5_{\mu 5}-\Gamma ^
\mu _{55}\Gamma ^\nu _{\nu \mu };
\eqry
and
\bqry
^{(5)}R & = &  ^{(4)}R+g^{\mu \nu }\left( \frac{\partial \Gamma ^5_
{\mu 5}}{\partial x^\nu }+\Gamma ^5_{\mu 5}\Gamma ^5_{\nu 5}-\Gamma ^
\rho _{\mu \nu }\Gamma ^5_{\rho 5}\right)  \NL
 &  & +g^{55}\left( -\frac{\partial \Gamma ^\mu _{55}}
{\partial x^\mu }+\Gamma ^\mu _{55}\Gamma ^5_{\mu 5}-\Gamma ^
\mu _{55}\Gamma ^\nu _{\nu \mu }\right) ,
\eqry
where we have dropped the indices on the left of the Christoffel symbol.

The D-dimensional gravitational field equations read $$^{(D)}R_{mn}-
\frac{1}{2}g_{mn}\;^{(D)}R=8\pi G\;^{(D)}T_{mn},\;\;\;m,n=0,1,
\ldots ,D-1.$$ Contracting with $g^{mn}$ gives $$\;^{(D)}R=-\frac{16
\pi G}{D-2}\;^{(D)}T^k_k,$$ and therefore, the field equations can be
rewritten as $$^{(D)}R_{mn}=8\pi G\left( \;^{(D)}T_{mn}-\frac{1}{D-2}
g_{mn}\;^{(D)}T^k_k\right) .$$ Thus, the 5D gravitational field
equations read
\beq
^{(5)}R_{\alpha \beta }=8\pi G\left( ^{(5)}T_{\alpha \beta }-\frac{1}{3}
g_{\alpha \beta }\;^{(5)}T^\lambda _\lambda \right) ,
\eeq
and can be written in the following component form:
\bqry
^{(5)}R_{\mu \nu } & = & 8\pi G\left( ^{(5)}T_{\mu \nu }-\frac{1}{2}g_
{\mu \nu }\;^{(5)}T^\rho _\rho \right) +\frac{4\pi G}{3}g_{\mu \nu }
\left( \;^{(5)}T_\rho ^\rho -2\;^{(5)}T^5_5\right) , \NL
^{(5)}R_{55} & = & \frac{8\pi G}{3}g_{55}\left( \;^{(5)}T^\rho _\rho -
2\;^{(5)}T^5_5\right) ,\;\;\; ^{(5)}R_{\mu 5}\;=\;0,\;\;\;\rho =0,1,2,3.
\eqry
The natural requirement on the 5D theory we are dealing with is to
contain the Einstein 4D theory of gravitation. Therefore, the 5D field
equations (A.6) should reduce to the Einstein 4D field equations in the
standard case. Such a case is $g_{55}={\rm const},$ i.e., no curvature
associated
with $\tau $ direction. In this case we have $$\Gamma ^5_{\mu 5}=\Gamma ^
\mu _{55}=0,\;\;\;^{(5)}R_{\mu \nu }=\;^{(4)}R_{\mu \nu },\;\;\;^{(5)}R_
{55}=0,\;\;\;^{(5)}R=\;^{(4)}R$$ and
\bqry
^{(4)}R_{\mu \nu } & = & 8\pi G\left( ^{(5)}T_{\mu \nu }-\frac{1}{2}
g_{\mu \nu }\;^{(5)}T^\rho _\rho \right) , \NL
^{(5)}T_{\mu 5} & = & 0,\;\;\;2\;^{(5)}T^5_5-\;^{(5)}T_\rho ^\rho \;=\;0.
\eqry
We require that (A.7) are exactly the Einstein equations; then \cite{Ma}
\beq
^{(5)}T_{\mu \nu }=\;^{(4)}T_{\mu \nu },\;\;\;\;^{(5)}T_{\mu 5}=0,\;\;\;
\;^{(5)}T^5_5=\frac{1}{2}\;^{(4)}T_\rho ^\rho .
\eeq
The quantity $^{(5)}T^5_5=\frac{1}{2}\;^{(4)}T^\rho _\rho $ was
calculated in ref. \cite{therm}:
\beq
^{(5)}T^5_5\equiv p_5=\sigma \mu _K\kappa ,
\eeq
where $\kappa $ is the density of the generalized Hamiltonian per unit
comoving three-volume (actually associated with the density of the
variable mass squared), $\mu _K$ is the mass potential in relativistic
ensemble \cite{HSP}, and $\sigma =\pm 1,$ according to invariance group
of an extended $(x^\mu ,\tau )$ manifold which could be $O(3,2)$ or
$O(4,1).$

Thus, the 5D gravitational field equations (A.5) with the source term
\beq
^{(5)}T_{\alpha \beta }=\left( \;^{(4)}T_{\mu \nu },\;p_5\right)
\eeq
reduce to the 4D Einstein equations in the assumption of no curvature
in $\tau $ direction.

As found in \cite{therm},
\beq
\mu _K\kappa \simeq \left\{ \begin{array}{ll}
p, & T\rightarrow \infty , \\
0, & T\rightarrow \;0.
\end{array} \right.
\eeq
Therefore, as $T\rightarrow 0,$
\beq
^{(5)}T_{\alpha \beta }=\left( \;^{(4)}T_{\mu \nu },0\right) ,
\eeq
where
\beq
^{(4)}T_{\mu \nu }=(p+\rho )u_\mu u_\nu -pg_{\mu \nu },\;\;\;u^\rho u_
\rho =1
\eeq
is the 4D energy-momentum tensor; \\ as $T\rightarrow \infty ,$ $$^{(5)}T
_{\alpha \beta }=\left( \;^{(4)}T_{\mu \nu },\sigma p\right) ,$$ i.e.,
\beq
^{(5)}T_{\alpha \beta }=(p+\rho )u_\alpha u_\beta -pg_{\alpha \beta },\;
\;\;u^\lambda u_\lambda =1,
\eeq
$$g_{\alpha \beta }=(g_{\mu \nu },\;\sigma ),\;\;\;\sigma =\pm 1,$$
represents the generalized 5D energy-momentum tensor.

The 5D geodesic equations
\beq
\frac{d^2x^\alpha }{ds^2}+\;^{(5)}\Gamma ^\alpha _{\beta \gamma }\frac{
dx^\beta }{ds}\frac{dx^\gamma }{ds}=0
\eeq
reduce, via (A.2), to
\beq
\frac{d^2x^\mu }{ds^2}+\;^{(5)}\Gamma ^\mu _{\nu \rho }\frac{dx^\nu }{ds}
\frac{dx^\rho }{ds}+\;^{(5)}\Gamma ^\mu _{55}\left( \frac{d\tau }{ds}
\right) ^2=0,
\eeq
\beq
\frac{d^2\tau }{ds^2}+2\;^{(5)}\Gamma ^5_{\mu 5}\frac{d\tau }{ds}
\frac{dx^\mu }{ds}=0.
\eeq
In the standard case $g_{55}=$const, $^{(5)}\Gamma ^5_{\mu 5}=\;^{(5)}
\Gamma ^\mu _{55}=0,$ and Eqs. (A.16),(A.17) take on the form
\beq
\frac{d^2x^\mu }{ds^2}+\;^{(5)}\Gamma ^\mu _{\nu \rho }\frac{dx^\nu }{ds}
\frac{dx^\rho }{ds}=0,
\eeq
\beq
\frac{d^2\tau }{ds^2}=0.
\eeq
It then follows that $\tau =as+b,\;\;a,b={\rm const,}\;\;a\neq 0,$ and
therefore, (A.18) reduces (since $^{(5)}\Gamma ^\mu _{\nu \rho }=\;^{(4)}
\Gamma ^\mu _{\nu \rho })$ to
\beq
\frac{d^2x^\mu }{d\tau ^2}+\;^{(4)}\Gamma ^\mu _{\nu \rho }
\frac{dx^\nu }{d\tau }\frac{dx^\rho }{d\tau }=0,
\eeq
the standard 4D geodesic equations for the metric
\beq
d\tau ^2=g_{\mu \nu }dx^\mu dx^\nu .
\eeq


\begin{thebibliography}{9}
\bibitem{hor} We assume that the phase transition occurs at $T\sim 10^
{12}$ K, which corresponds to cosmic time $t\sim 10^{-5}\;c$ (see ref.
[18]).
\end{thebibliography}
\end{document}